\documentclass[10pt,journal,compsoc]{IEEEtran}
\newif\ifpeerreview

\peerreviewfalse

\usepackage[nocompress]{cite}
\usepackage{url}
\usepackage{amsmath,amssymb,graphicx}

\usepackage{lipsum} 
\usepackage{mathrsfs,amsmath}
\usepackage{comment}
\usepackage{scalerel}
\usepackage[overload]{empheq}
\usepackage{nicefrac,xfrac}
\usepackage{mathtools}
\usepackage{gensymb}
\usepackage{soul}
\usepackage{textgreek}

\usepackage{amsmath}
\usepackage{bm}
\usepackage{algorithm}
\usepackage{algpseudocode}
\usepackage{enumitem}
\usepackage{amsmath}
\usepackage{setspace}
\usepackage{xcolor}
\usepackage{gensymb}
\usepackage{multicol,lipsum}

\usepackage{mathtools}
\usepackage{cuted}

\providecommand{\vrr}{\mathbf{r}}
\providecommand{\vu}{\mathbf{u}}

\providecommand{\vv}{\mathbf{v}}

\DeclarePairedDelimiter\abs{\lvert}{\rvert}%
\DeclarePairedDelimiter\norm{\lVert}{\rVert}%

\DeclareMathOperator*{\argmin}{arg\,min}%

\usepackage[switch]{lineno}

\newcommand{\paperID}{39}

\title{Tensorial tomographic differential phase-contrast microscopy}

\author{Shiqi~Xu,~Xiang~Dai,~Xi~Yang,~Kevin~C.~Zhou,~Kanghyun~Kim,~Vinayak~Pathak,~Carolyn~Glass~        and~Roarke~Horstmeyer 
\IEEEcompsocitemizethanks{\IEEEcompsocthanksitem S. Xu, X. Yang, K. Zhou, K. Kim, V. Pathak, and R. Horstmeyer are with the Department
of Biomedical Engineering, Duke University, Durham,
NC, 27708.\protect\\
E-mail: shiqi.xu@duke.edu
\IEEEcompsocthanksitem X. Dai is with the Department
of Electrical and Computer Engineering, University of California San Diego, San Diego, CA, 92161.\IEEEcompsocthanksitem C. Glass is with Duke University Medical Center, Durham, NC 27708.}
}

\begin{document}

\IEEEtitleabstractindextext{%
\begin{abstract}
We report \emph{\textbf{T}ensorial \textbf{T}omographic \textbf{D}ifferential \textbf{P}hase-\textbf{C}ontrast microscopy (T$^2$DPC)}, a quantitative label-free tomographic imaging method for simultaneous measurement of phase and anisotropy. T$^2$DPC extends differential phase-contrast microscopy, a quantitative phase imaging technique, to highlight the vectorial nature of light. The method solves for permittivity tensor of anisotropic samples from intensity measurements acquired with a standard microscope equipped with an LED matrix, a circular polarizer, and a polarization-sensitive camera. We demonstrate accurate volumetric reconstructions of refractive index, birefringence, and orientation for various validation samples, and show that the reconstructed polarization structures of a biological specimen are predictive of pathology. 
\end{abstract}

}

\maketitle
\thispagestyle{empty}

\IEEEraisesectionheading{
  \section{Introduction}\label{sec:introduction}
}
%
%
%
%
\IEEEPARstart{Q}{uantitative} phase imaging (QPI) is a new emerging class of label-free microscopy method to measure endogenous contrast of cell and tissue by turning specimen-induced phase or polarization changes into images. Due to its low phototoxicity and no photobleaching, QPI has become a valuable tool for studying biology, chemistry, and medicine~\cite{park2018quantitative}. The fine anisotropic structures of biological samples are characteristic of their molecular arrangements, such as cell membrane and axon bundles, which can now be detected with high sensitivity at scales smaller than the wavelength of light\cite{zhanghao2020high,lu2020single}. As such, polarization-sensitive microscopy is widely used in pathology~\cite{he2021polarisation}, developmental biology~\cite{biopol}, and mineralogy~\cite{panwar2020review}, especially in combination with phase microscopy techniques\cite{guo2020revealing}. 

In the past, numerous 2D label-free polarization microscopy techniques have been invented, from classical qualitative methods such as differential interference contrast (DIC) microscopy~\cite{nomarski1955differential} and polarized light microscopy~\cite{schmidt1924bausteine,inoue1953polarization}, to more contemporary polarization microscopy in transmission mode that reports quantitative retardance and 3D orientation of optical path~\cite{oldenbourg2013polarized,mehta2013polarized,spiesz2011quantitative}. Diffractive polarization microscopy in particular has attracted many attentions recently. In general there are two categories of methods to achieve quantitative polarization microscopic imaging in combination with diffractive optics: holographic measurements and computational illumination. The first category typically employs interferometry and hence requires highly coherent laser illumination~\cite{jiao2020real,shin2018reference,ge2021single,liu2020deep}. While those methods can image quantitative Jones matrix at high-speed, holographic systems still remain challenging to align, and coherent speckle artifacts maybe present to decline image quality.

Computational illumination methods, on the other hands, very often use partially coherent light sources, either from a programmable LED array or LCD display, to illuminate the sample and create a data stack, from which the phase and anisotropy images are then computationally recovered~\cite{song2020ptychography,song2021large,guo2020revealing,dai2022quantitative,hur2021polarization}. This has the flexibility for creating not only angular diversity in the illumination to enable high spatial-bandwidth product images~\cite{song2021large,dai2022quantitative}, but also polarization diversity to further incorporate the depolarization effects~\cite{guo2020revealing}, for instance. 

To image thicker samples, three-dimensional imaging method are desired to obtain a depth-resolved representation. One well-established technique is polarization sensitive optical coherence tomography~\cite{de2017polarization}, which creates cross-sectional images for tissues by point-scanning.  Recently reported polarization-sensitive diffraction tomography approaches~\cite{van2020polarization,saba2021polarization} create volumetric reconstructions of polarization properties that highlights biologically specific structures in model organisms~\cite{van2020polarization}. While often impressive, these methods are usually based on Michelson interferometer setup that requires precise mechanics to either rotate the sample or steer the beam illuminated from different angles, which are challenging to implement. Alternate approaches use high numerical aperture (NA) objective to scan through the semi-transparent samples\cite{yeh2021upti}, creating "confocal-like" depth sections of the entire mouse brain. However, such methods require expensive oil-immersion lenses and additional high-quality illumination units\cite{yeh2021upti}.

\begin{figure*}[!t]
\centering
\includegraphics[width=14cm]{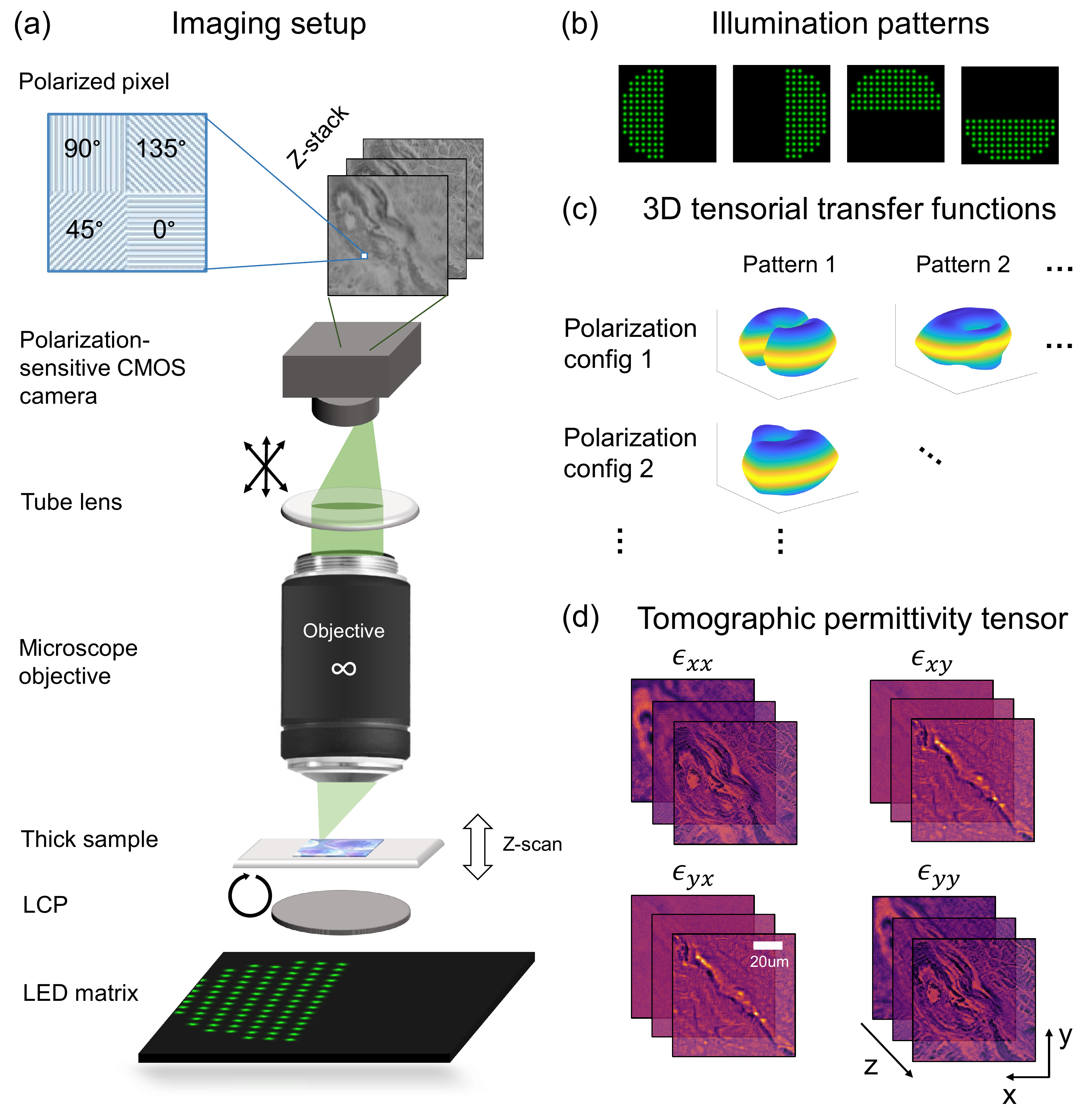}
\caption{Workflow of T$^2$DPC. (a) An illustration of the experimental setup. Light from LED array is polarized with a left circular polarizer (LCP) before illuminating the sample. After interacting with the permittivity tensor of the sample, light propagates through a 4f imaging system. A camera with polarization sensitive sensor is place at back focal plane of the tube lens to record $0\degree, 45\degree, 90\degree, 135\degree$ oriented components of the vectorial light. (b) displays the illumination patterns we use. (c) plots a few representative transfer functions for different polarization measurement configuration under different illumination patterns. (d) shows reconstructed tomographic permittivity tensor of a human heart tissue sample.}
\label{fig1::overview}
\end{figure*}

To overcome the challenges of existing techniques, we propose T$^2$DPC, a new 3D differential phase contrast-based polarization sensitive tomographic microscope technique. Differential phase contrast microscopy (DPC) is a low-cost label-free method that can create quantitative tomographic reconstruction of refractive index by capturing a focal stack of the sample of interest illuminated with partially coherent light~\cite{chen20163d}. Based on DPC principles, we demonstrate how we can outfit a standard microscope with an programable LED matrix, a circular polarizer, and a polarization sensitive camera to solve for the permittivity tensor maps of anisotropic samples. In the rest of this article, we first derive a forward model describing how vectorial light propagation through tensorial samples for our experimental setup. We then evaluate our method by analyzing tomographic reconstructions of refractive index, birefringence, and orientation for various calibration samples, and demonstrate that the reconstructed polarization properties of a biological specimen is predictive of pathology.

\section{Proposed Method}
Here we describe principle and experimental implementation of T$^2$DPC.
\subsection{Experimental setup}
The experimental setup is illustrated in Fig.\ref{fig1::overview}(a), whose backbone is an inverted microscope. The sample was illuminated from the bottom by an LED matrix with $25\times25$ addressable pixels (WS2812B-2020) and a pitch of 3.125 mm. The DPC illumination patterns are depicted in 
Fig.\ref{fig1::overview}(b). We note that there are more advanced illumination patterns, such as ring and gradient patterns, for instance, for improved DPC contrast and robustness\cite{fan2019optimal,cao2022self}. Here, we use semicircular illumination patterns as a first demonstration, as they are standard\cite{chen20163d,cao2022self} and give better signal-to-noise ratio (SNR) due to higher illumination
radiance. The maximum illumination angle is chosen to match the NA of the objective. To generate a polarized illumination source, we placed a left circular polarizer (Edmund CP42HE) between the LED array and sample. The anisotropic specimen is placed on a mechanical stage (actuator from Thorlabs Z825b) to translate the sample along the optical axis to create a focal stack. After getting diffracted by the sample, light propagates through a 4f system consisting of an objective (Olympus PLN; 20x, 0.4NA for validation samples, and 10x, 0.25NA for biological samples) and a 180-mm tube lens (Thorlabs AC508). An image is formed in the back focal plane of the tube lens, and captured with a polarization sensitive sensor array (Sony IMX250MZR), which detects intensity of four linearly polarized components of the light field. The polarized CMOS sensor achieves this by placing micro-lens and wire-grid linear polarizer oriented at $0\degree, 45\degree, 90\degree,$ and $ 135\degree$ in front of each individual camera pixels~\cite{maruyama20183}, as illustrated in Fig.\ref{fig1::overview}(a). Each pixel has a pixel size (pitch) of 3.45 \textmu{m}; hence, a two-by-two ``super-pixel'' contains all four polarization measurements, with an effective pitch of 6.9 \textmu{m}. A micro-controller (ARM Cortex-M3) along with a voltage level shifter (Todiys SN74AHCT) are used to control the LED illumination and hardware trigger the motorized actuator.

\subsection{Principle of T$^2$DPC}
\subsubsection{Notation}
Throughout this article, we use vector symbol $\vec{\cdot}$  and matrix symbol $\bar{\bar{\cdot}}$ to represent interaction between polarized light and specimen tensors. Further, we use \textit{Mathematical Script} font with curly brackets to describe spatial operations, such as $\mathcal{F}\{\cdot\}$, which denotes Fourier transform. Bold letters in lower case are used to describe vectors in frequency ($\vu$) or space ($\vrr$). Further, we use $\tilde{\cdot}$ to indicate frequency domain counterparts of variables first defined in space domain.
\subsubsection{Light propagation}
\label{sec::physics}
A sample's vectorial optical property can be described by its $3\times3$ permittivity matrix~\cite{born2013principles} 
\begin{equation}
\bar{\bar{\epsilon}}=
    \begin{bmatrix}
\epsilon_{xx}(\vrr) & \epsilon_{xy}(\vrr) & \epsilon_{xz}(\vrr)\\
\epsilon_{yx}(\vrr) & \epsilon_{yy}(\vrr) &
\epsilon_{yz}(\vrr)\\
\epsilon_{zx}(\vrr) & \epsilon_{zy}(\vrr) &
\epsilon_{zz}(\vrr)
\end{bmatrix},
\end{equation}
where $\vrr=(x, y, z)$ is the voxel position in three dimension. Under the first Born approximation~\cite{yeh2021upti,saba2021polarization}, the relation between scattered light field $\vec{E}^s$ and illumination field $\vec{E}^0 $ can be written as 
\begin{equation}
     \vec{E}^s(\vrr)=
     \iiint\bar{\bar{G}}(\vrr-\vrr')\bar{\bar{V}}(\vrr')
     \vec{E}^0(\vrr')d\vrr',
\end{equation}
where $\bar{\bar{V}}(\vrr') = \bar{\bar{\epsilon}} - \bar{\bar{\epsilon}}_0$ is the sample scattering potential matrix, $\bar{\bar{\epsilon}}_0$ is the permitivity tensor of the background medium, and $\bar{\bar{G}}(\vrr)$ is the dyadic Green's tensor~\cite{yaghjian1978direct}. Assuming each quasi-monochromatic LED source with wavelength $\lambda$ generates a field at the sample plane, with a lateral frequency $\vu'=(u_x',u_y')$ and axial frequency $\eta=u_z'$; $\abs{\vu}^2+\eta^2=\lambda^{-2}$. After the vectorial electric field is diffracted by the sample, the optical field propagates through the imaging system, which can be modeled as a low-pass filter with a pupil Jones matrix $\bar{\bar{P}}(\vu)$ in the frequency domain~\cite{dai2022quantitative}. Hence, if we denote the illumination as $\vec{E}^0(\vrr,\vu')$, the detected intensity on the image plane with $l^{th}$ analyzer (linear polarizers at $\{0\degree,45\degree,90\degree,135\degree\}$) with single LED illumination from angle $\vu'$ can be written as~\cite{dai2022quantitative,ferrand2018quantitative}
\begin{equation}
        I^{l}(\vrr,\vu')=\abs{{\vec{a}_l}^T\mathcal{F}_{2d}^{-1}\bar{\bar{P}}(\vu)\mathcal{F}_{2d}(\vec{E}^s(\vrr,\vu')+\vec{E}^0(\vrr,\vu'))}^2.
\label{eq:forward_general}
\end{equation}
$\vec{a}_l$ is the Jones vector for $l^{th}$ analyzer. For a linear polarizer oriented at $\alpha$, $\vec{a}=[\cos{\alpha},\sin{\alpha}]^T$~\cite{ferrand2015ptychography,dai2022quantitative}. The intuition behind this is that after linear analyzer, the vectorial field reduces to a scalar pointing along one polarization direction~\cite{dai2022quantitative}.

For DPC measurements, we assume different LEDs are mutually incoherent. Thus, the measured intensity from each pattern illumination is incoherent sum of intensities from each individual point sources. To approximate this summation with an integral, the measured intensity from $n^{th}$ illumination pattern $S^n(\vu_0)$ is 
\begin{equation}
    I^{l,n}(\vrr)=\iint{S^n(\vu')}I^{l}(\vrr,\vu')d\vu'.
\end{equation}

\subsubsection{Forward model and inverse problem}
In our first demonstration of T$^2$DPC, we make a few approximations to the model described in Section \ref{sec::physics} to obtain a less ill-posed inverse problem for image reconstruction with the proposed experimental measurement strategy.

Following Saba \textit{et.al.}~\cite{saba2021polarization}, we first make a paraxial approximation. Assuming weak polarization along optical axis of the illumination, and negligible interaction between traverse and axial polarization caused by the sample, we can approximate the $3\times3$ permitivity matrix with a $2\times2$ one, 
\begin{equation}
\bar{\bar{\epsilon}}=
    \begin{bmatrix}
\epsilon_{xx}(\vrr) & \epsilon_{xy}(\vrr)\\
\epsilon_{yx}(\vrr) & \epsilon_{yy}(\vrr)
\end{bmatrix}.
\label{eq:permit_define}
\end{equation}
Although this approximation can be inaccurate for certain crystals illuminated at high angle~\cite{saleh2019fundamentals}, a finite element analysis-based study showed that it is accurate up to a $25\degree$ illumination angle~\cite{saba2021polarization}. In addition, we assume the background media is isotropic, which has a diagonal permittivity tensor $\bar{\bar{\epsilon}}_0=\epsilon_0\mathbb{I}$. This also simplifies the Green's tensor to a diagonal matrix with same component for each polarization~\cite{saba2021polarization}
\begin{equation}
    \bar{\bar{G}}(\vrr,\vrr')=\begin{bmatrix}
      G(\vrr,\vrr') & 0  \\
      0 & G(\vrr,\vrr'),
     \end{bmatrix}
\end{equation}
in which 
\begin{equation}
\label{eq:gfuc_real}
    G(\vrr,\vrr')=\frac{e^{jk_0\abs{\vrr-\vrr'}}}{4\pi\abs{\vrr-\vrr'}}
\end{equation}
is the same as the scalar Green's function.

We further assume the sample we image is homogeneous. That is, there exist a {principal axis}, where the permitivity matrix becomes diagonal\cite{born2013principles,saleh2019fundamentals,jones1941new}. This is widely assumed for various crystals and biological samples\cite{chipman2018polarized,mehta2013polarized}. Under this assumption, the permittivity matrix is symmetric, and can be decomposed into~\cite{chipman2018polarized}, \begin{equation}
\label{rot_mtx}
\bar{\bar{\epsilon}}=
    \begin{bmatrix}
\cos\theta & \sin\theta\\
-\sin\theta & \cos\theta
\end{bmatrix} \begin{bmatrix}
\epsilon_e & 0\\
0 & \epsilon_o
\end{bmatrix} \begin{bmatrix}
\cos\theta & -\sin\theta\\
\sin\theta & \cos\theta
\end{bmatrix}.
\end{equation}
Here $\epsilon_o$ and $\epsilon_e$ are permittivity along ordinary and extraordinary axis, and $\theta$ is the orientation to the extraordinary axis, or the slow axis (extraordinary axis has a higher refractive index). Following the convention~\cite{jones1941new}, we rename elements in Eq.\ref{eq:def_eps} as
\begin{subequations}
\label{eq:def_eps}
    \begin{empheq}[left={\empheqlbrace\,}]{align}
     \epsilon_1 &= \epsilon_{xx}\\
     \epsilon_2 &= \epsilon_{yy}\\
     \epsilon_3 &= \epsilon_{xy}=\epsilon_{yx},
    \end{empheq}
\end{subequations}
With left circularly polarized illumination, the scattering potential along $x-y$ polarization is
\begin{equation}
     \begin{bmatrix}
V_1 & V_3\\
V_3 & V_2
\end{bmatrix}\begin{bmatrix}1\\j\end{bmatrix}=\begin{bmatrix}V_1+jV_3\\V_3+jV_2\end{bmatrix},
\end{equation}
where
\begin{subequations}
\label{eq:def_v}
    \begin{empheq}[left={\empheqlbrace\,}]{align}
     V_1 &= 4\pi{k}_0^2(\epsilon_1-\epsilon_0)\\
     V_2 &= 4\pi{k}_0^2(\epsilon_2-\epsilon_0)\\
     V_3 &= 4\pi{k}_0^2\epsilon_3,
    \end{empheq}
\end{subequations}
and $k_0$ is the wavenumber of the isotropic background medium.

\begin{figure*}[!t]
\centering
\includegraphics[width=14cm]{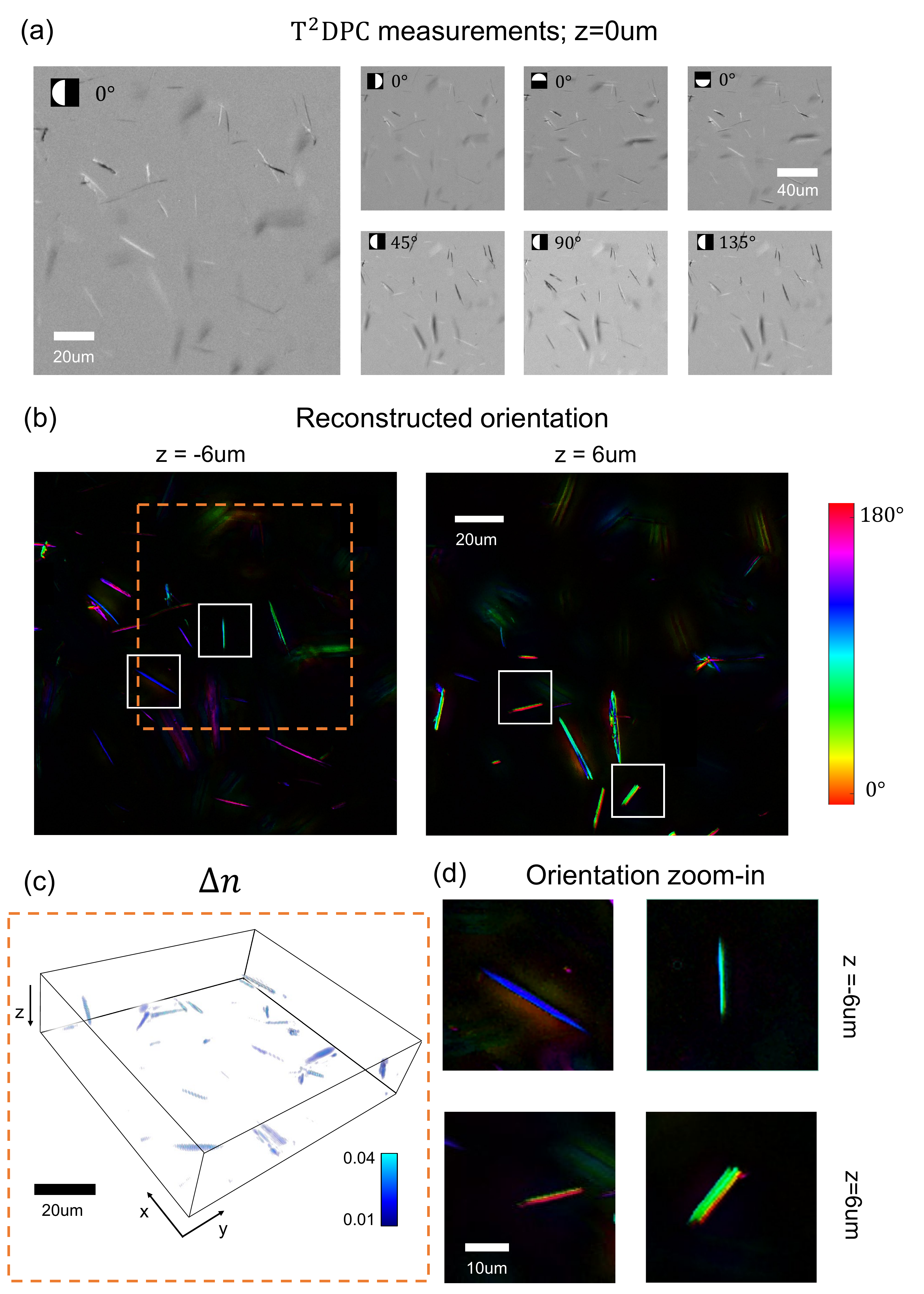}
\caption{Reconstruction of a two-layer monosodium urate (MSU) crystal sample. (a) shows a few representative images of raw measurements, when focused in the middle of the sample. First row shows images captured with different illumination patterns. Second row shows images captured with analyzers oriented at different directions. (b) plots reconstruction slices of MSU sample at two unique z-positions. The reconstructed orientation for each pixel is coded with color. (c) A 3D view of reconstructed birefringence of the sample, with its field-of-view labeled in (b). (d) Zoom-in for the orientation reconstructions regions labeled in (b)}
\label{fig2:msu}
\end{figure*}

\begin{figure*}[!t]
\centering
\includegraphics[width=14cm]{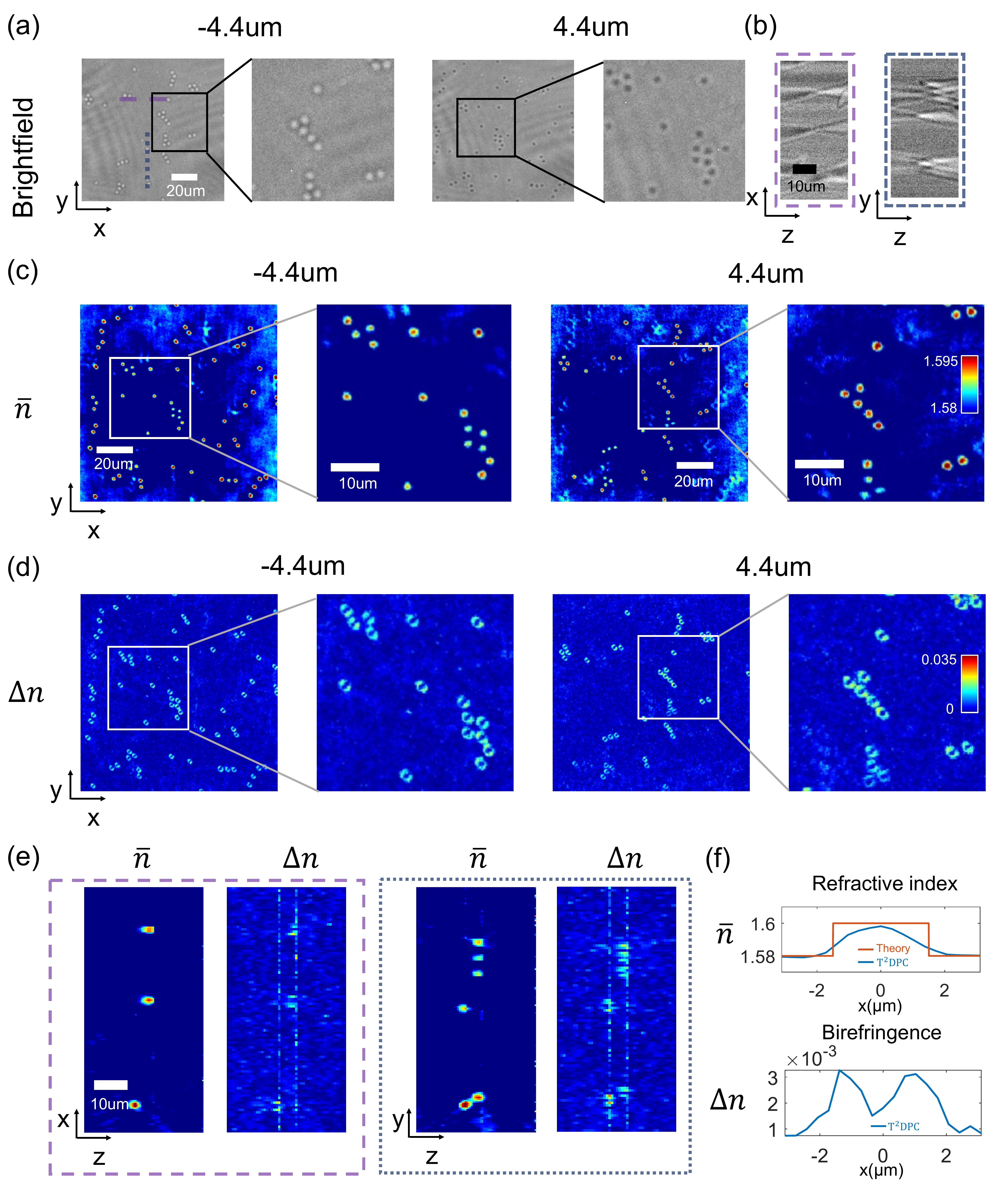}
\caption{Validation reconstructions from a two-layer sample made with 3-\textmu m polystrene microspheres immersed in $n=1.58$ oil. (a) shows brightfield images at two different z positions, serving as validation images. Note that for a transparent sample, the contrast is low when in focus. (b) display x-z and y-z cross sections of brightfield focal stack measurement at two positions, labeled in (a).  (c) and (d) display two reconstructed slices for refractive index and birefringence at different depths. Each column in (c-d) share the same scale bar. (e) plots x-z and y-z cross sections of the reconstruction volume at two positions, labeled in (a). (f) plots the averaged line profile of ten microspheres along x-axis.}
\label{fig3::beads}
\end{figure*}

\begin{figure*}[!t]
\centering
\includegraphics[width=13.cm]{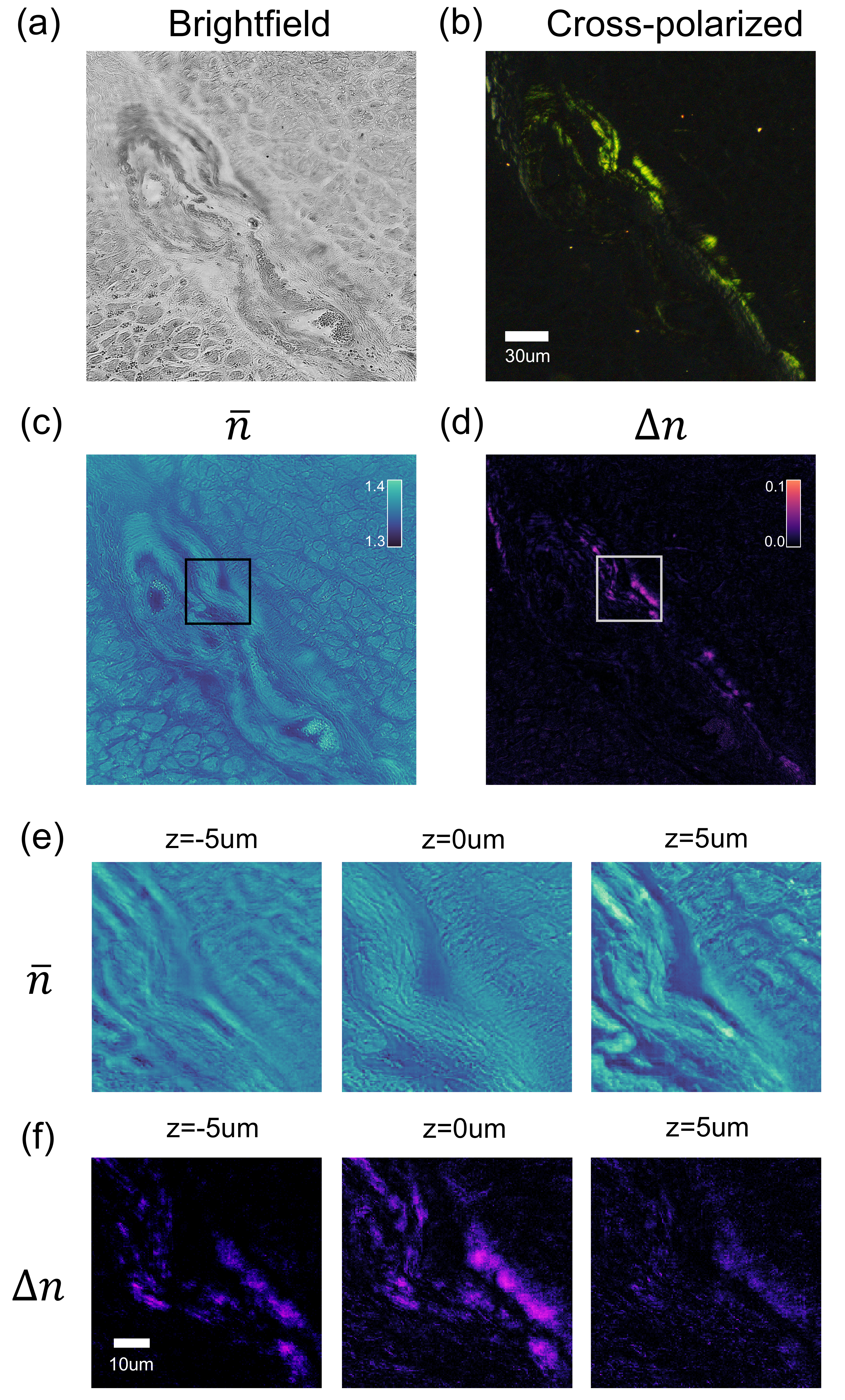}
\caption{Images of congo red stained cardiac tissue sample. (a) is a brightfield image of the sample. (b) is a color image taken with color sensor, when generator and analyzer are cross-polarized, serving as a validation image. (c) presents a central slice of the 3D reconstructed refractive index of the sample. (d) highlights a central slice of the reconstructed 3D birefringence of the sample. (e-f) display reconstructed slices at different z-positions for the region labelled in (c) and (d). (c) and (e) share the same colormap. (d) and (f) share the same colormap.}
\label{fig4::congored}
\end{figure*}

Finally, following Chen \textit{et.al.}~\cite{chen20163d}, we make two more approximations to the illumination and scattering process. First we assume the illumination at sample plane from each LED is a plane wave, which is widely assumed in LED-based computational microscopy\cite{konda2020fourier}. Second, we apply a weak object approximation, which neglect the second order scattering term~\cite{li2019high,chen20163d,ayoub20213d}, and linearize the model. Further, since high-quality scientific grade objectives (Olympus PLN) are used in this experiment, we disregard pupil aberration in this first demonstration, and model the pupil it as a low-pass filter $P(\vu)$ with cut-off frequency equivalent to the NA of the objective lens~\cite{chen20163d}. Joint reconstruction of pupil Jones matrix for anisotropic aberration correction is left to future work~\cite{dai2022quantitative}. The T$^2$DPC forward model we use can then be expressed as
\begin{align}
\begin{split}
     \tilde{I}^{l,m}(\vu,\eta)=\tilde{I}_0^{l,m}(\vu,\eta)+{H}_{Re}^m&(\vu,\eta)\cdot\tilde{V}_{Re}^l(\vu,\eta)\\&+{H}_{Im}^m(\vu,\eta)\cdot\tilde{V}_{Im}^l(\vu,\eta),
\end{split}
\end{align}
where $\tilde{I}^{l,n}(\vu,\eta)$ is 3D Fourier transform of the measured focal stack, and 3D Fourier transform of the DC term represents the background intensity illuminated with $n^{th}$ pattern and analyzed with $l^{th}$ analyzer; $l\in\{0\degree,45\degree,90\degree,135\degree\}$. $\tilde{I}_0^{l,m}(\vu,\eta)$ is  $\tilde{V}_{Re}^l(\vu,\eta)$ and $\tilde{V}_{Im}^l(\vu,\eta)$ are 3D Fourier transforms of ${V}_{Re}^l(\vrr)$ and ${V}_{Im}^l(\vrr)$, respectively, and are related to Eq. \ref{eq:def_v} via
\begin{subequations}
\label{eq:def_vdegree}
    {\begin{empheq}[left={\empheqlbrace\,}]{align}
    &{V}_{Re}^{0\degree}(\vrr) + j{V}_{Im}^{0\degree}(\vrr) =  V_1(\vrr)+jV_3(\vrr)\\
     \begin{split} &{V}_{Re}^{45\degree}(\vrr) + j{V}_{Im}^{45\degree}(\vrr) = 
         \frac{\sqrt{2}}{2}(V_1(\vrr)+V_3(\vrr))\\&\quad\quad\quad\quad\quad\quad\quad\quad\quad\quad+j\frac{\sqrt{2}}{2}(V_2(\vrr)+V_3(\vrr))
     \end{split}\\
     &{V}_{Re}^{90\degree}(\vrr) + j{V}_{Im}^{90\degree}(\vrr) = V_3(\vrr)+jV_2(\vrr)\\
          \begin{split} &{V}_{Re}^{135\degree}(\vrr) + j{V}_{Im}^{135\degree}(\vrr)=  \frac{\sqrt{2}}{2}(-V_1(\vrr)+V_3(\vrr))\\&\quad\quad\quad\quad\quad\quad\quad\quad\quad\quad+j\frac{\sqrt{2}}{2}(V_2(\vrr)-V_3(\vrr)),\end{split}
    \end{empheq}}
\end{subequations}
which are the scattering potential components corresponding to each analyzer angles. Note that $V_1(\vrr)$, $V_2(\vrr)$, and $V_3(\vrr)$ can have imaginary parts; thus, the above equation does not necessarily suggest $ \tilde{V}_{Re}(\vrr)=V_1(\vrr)$, for instance. 
\begin{equation}
    {\small\begin{split}
    &H_{Re}^m(\vu, \eta)=\iint{S}^m(\vu')\cdot\big[P^*(-\vu)\cdot\tilde{G}(-\vu+\vu_0, -\eta+\eta_0)\\&\cdot{P}(-\vu+\vu_0)+{P}(-\vu)\cdot\tilde{G}^*(-\vu-\vu_0, -\eta-\eta_0)\\&\cdot{P}^*(-\vu-\vu_0)\big]d^2\vu'
    \end{split}}
\end{equation}
and $H_{Im}^{m}=j\cdot{H}_{Re}^m$ are the transfer functions for the real and imaginary part of the scattering potential when illuminated with $m^{th}$ pattern ${S}^m(\vu')$, respectively~\cite{chen20163d,streibl1985three}. $\tilde{G}(\vu,\eta)$ is the Fourier transform of Green's function $G(\vrr)$ in Eq.(\ref{eq:gfuc_real}). For concise expression, we now denote $\mathbf{v}\in\mathbb{C}^{N\times M\times Q\times3}$ as a vector representation of $V_1, V_2, V_3$, with $N,M,Q$ are width, height, and depth of the 3D sample, respectively. {The sizes of reconstructed digital images in Fig.~\ref{fig2:msu}-\ref{fig4::congored} are $512\times512\times100\times3$, $400\times400\times100\times3$, and $1024\times1024\times80\times3$, with physical voxel sizes of $0.345\times0.345\times0.8\mu\text{m}^3$, $0.345\times0.345\times0.8\mu\text{m}^3$, and $0.69\times0.69\times1.0\mu\text{m}^3$, respectively. The measurements have the same digital image size as the reconstruction, with 20x magnifications used in Fig.~\ref{fig2:msu}-\ref{fig3::beads} and a 10x magnification for Fig.~\ref{fig4::congored}.} We further introduce the operator $\mathcal{A}_{\mathbb{U}}(\cdot)$ as the forward model operator for all the polarization configurations, where $\mathbb{U}$ denotes the illumination angular support: $\mathbb{U}=[\vu_{min},\vu_{max}]\cap[\eta_{min},\eta_{max}]$. To reconstruct the permittivity matrix, the inverse problem is formulated as
\begin{equation}
    \mathbf{v} = \argmin_{\mathbf{v}}\mathcal{L}(\mathbf{v}),
\end{equation}
with the loss function
\begin{equation}
    \mathcal{L}_{\mathbb{U}}(\mathbf{v}) = \sum_{\{\vu,\eta\}\in\mathbb{U}}\norm{\mathcal{A}_{\mathbb{U}}(\mathbf{v})-{\tilde{I}}(\vu,\eta)}_2^2+\gamma\sum_i\text{tv}(\mathbf{v}).
\end{equation}

$\text{tv}(\cdot)$ is the isotropic total variation operator applied on $V_1, V_2, V_3$. $\rho$ is a regularization penalization parameter empirically set to be $1\times10^{-7}$ for all the experiments. The forward model is implemented in Pytorch~\cite{paszke2017automatic}, an auto-differentiation library, and the loss is minimized using the Adam optimizer~\cite{kingma2014adam}. Due to the memory limitation, we use stochastic gradients to approximate the full gradient,
\begin{equation}
    \frac{\partial}{\partial\mathbf{v}}\mathcal{L}_{\mathbb{U}}(\mathbf{v})\approx\frac{\partial}{\partial\mathbf{v}}\mathcal{L}_{\mathbb{U}_s}(\mathbf{v}),
\end{equation}
where $\mathbb{U}_s\subset\mathbb{U}$ is a subset of the support of $\mathcal{A}_{\mathbb{U}}(\vv)$ chosen randomly at each iteration.

\subsubsection{Deriving polarization properties from T$^2$DPC reconstruction}
To obtain polarization properties such as orientation and birefringence from T$^2$DPC reconstructions, we start by explicitly writing out the following relation described in Eq(\ref{rot_mtx}),
\begin{subequations}
\begin{empheq}[left={\empheqlbrace\,}]{align}
  \epsilon_1 &= \epsilon_e\cos{\theta}^2+\epsilon_o\sin{\theta}^2\\
    \epsilon_2 &= \epsilon_e\sin{\theta}^2+\epsilon_o\cos{\theta}^2\\
    \epsilon_3 &= (\epsilon_e-\epsilon_o)\sin{\theta}\cos{\theta},
    \end{empheq}
\end{subequations}
from which, we can compute
\begin{subequations}
\begin{empheq}[left={\empheqlbrace\,}]{align}
  \epsilon_1 + \epsilon_2 &= \epsilon_o+\epsilon_e\\
    \epsilon_1 - \epsilon_2 &= (\epsilon_e-\epsilon_o)\cos(2\theta)\\
    2\epsilon_3 &= (\epsilon_e-\epsilon_o)\sin(2\theta).
    \end{empheq}
\end{subequations}
Thus, we have
\begin{subequations}
\begin{empheq}[left={\empheqlbrace\,}]{align}
    \epsilon_0 &= \bar{\epsilon}-\nicefrac{1}{2}\Delta\epsilon \\
        \epsilon_e &= \bar{\epsilon}+\nicefrac{1}{2}\Delta\epsilon,
        \end{empheq}
\end{subequations}
where
\begin{subequations}
\begin{empheq}[left={\empheqlbrace\,}]{align}
    \Delta\epsilon&=\epsilon_e-\epsilon_o = \sqrt{( \epsilon_1 - \epsilon_2)^2+4\epsilon_3^2}\\
    \bar{\epsilon}&=\frac{\epsilon_1 + \epsilon_2}{2}.
        \end{empheq}
\end{subequations}

Then the refractive index along the ordinary and extraordinary axes can be computed as $n_{o,e}=\sqrt{\epsilon_{o,e}}$, along with the averaged refractive index ($\bar{n}=n_o+n_e$) and birefringence ($\Delta{n}=n_e-n_o$) reported in Fig. \ref{fig3::beads} and Fig. \ref{fig4::congored}. Finally, the orientation of the slow or extraordinary axis $\theta\in[0,\pi]$ reported in Fig. \ref{fig2:msu} can be derived using the relation
\begin{equation}
    \tan{2\theta}=\frac{2\epsilon_3}{\epsilon_1-\epsilon_2}.
\end{equation}
Hence, 

\begin{equation}
    \theta=\begin{cases}
		  \frac{1}{2}\arctan\nicefrac{2\epsilon_3}{\epsilon_1-\epsilon_2}, \;  &\text{if}\; \epsilon_1-\epsilon_2>0\\
           \frac{1}{2}\arctan\nicefrac{2\epsilon_3}{\epsilon_1-\epsilon_2}+\frac{\pi}{2}, \; &\text{otherwise.}
		 \end{cases}
\label{eq:ori}
\end{equation}
\section{Experimental Results}
To evaluate the performance of proposed method, we show reconstructions of a variety of calibration samples, validating different aspects of the proposed method using different calibration samples. Finally, we show reconstruction of a heart tissue sample containing amyloid, an indicator of a lethal heart disease called cardiac amyloidosis.
\subsection{Quantitative orientation measurement}
To validate the reconstructed orientation accuracy, we imaged two-layer monosodium urate (MSU) samples separated by 12 \textmu m. The focal stack is taken with 0.8\textmu{m} step size using an 0.4 NA objective. Figure \ref{fig2:msu}(a) shows the images taken when focused in middle of two layers. The first row shows representative images captured with the zero-degree analyzer pixels, and different illumination patterns. In contrast, the second row shows images when sample is illuminated with the same pattern, but analyzed with different linear polarizers. The pattern and analyzer orientation are labeled at top-left corners of each individual images. Figure \ref{fig2:msu}(b) shows reconstruction slices from two different z-positions. The spatially resolved orientation is computed using Eq.\ref{eq:ori}. To best visualize orientation results, we follow the convention~\cite{liu2020deep,song2021large,dai2022quantitative} to display the multidimensional polarization data using an HSV colormap, where value displays birefringence, orientation is coded in hue, and saturation is set to one. Zoom-ins of select regions are also presented in (d). These results suggest that the reconstructed orientations of line-shape MSU samples follow the structural directions of the sample, in agreement with prior works~\cite{liu2020deep,song2021large,dai2022quantitative}. Figure \ref{fig2:msu}(c) shows a 3D view of the reconstructed birefringence for the field-of-view (FOV) labelled in (b). {In addition, supplement Fig.1 presents 2D reconstructions of the MSU sample using a classic LC-PolScope method~\cite{oldenbourg2013polarized}. The same depth and zoom-in regions shown in Fig.2 are displayed here to highlight the superior depth sectioning ability of the proposed T$^2$DPC method. Although LC-PolScope can give accurate orientation and birefringence reconstructions from focused depth, artifacts are present in the images due to out-of-focus regions of the 3D sample.} 
\subsection{Validation of refractive index and birefringence}
In this section we show reconstruction results of a sample consisting of two layers of 3-\textmu{m} polystyrene microspheres (1.60 at 520 nm~\cite{sultanova2009dispersion}) immersed in 1.58-index oil. Similar to the configuration used to image MSU sample, we use 0.4NA objective lens, and 0.8\textmu{m} z-scan step size. Figure \ref{fig3::beads}(a) shows brightfield images captured when focused at different axial positions of the sample. Note that since polysterene spheres are transparent, and have a refractive index close to that of the immersion fluid, the beads almost disappear when they are in focus. These images confirm that the two layers of beads are separated by 8.8um. Figure \ref{fig3::beads}(c-d) presents reconstructed refractive index and birefringence for both layers. Edge birefringence is a well-established phenomenon\cite{oldenbourg1991analysis}, and we see expected ring-shaped birefringence reconstructions that match structures reported in prior literature\cite{yeh2021upti,dai2022quantitative}. Figure \ref{fig3::beads}(e) displays cross-sections of the reconstructed volume at two regions labelled in (a). We see the reconstructed two-layers are accurately separated by 8.8\textmu{m} with a better resolution comparing with the brightfield focal stack in (b). In addition, we see two line-shaped birefringence reconstructions. We suspect this is due to the edge birefringence at the boundary of oil and glass (both slide and cover glass). (f) plots a line profile of refractive index and birefringence reconstructions along x-axis, averaged from ten beads. {The reconstructed refractive index is slightly lower than the ground truth. The reconstructed birefrigence value matches results reported with other techniques ~\cite{dai2022quantitative} ($0.4$rad for a single layer of $10\mu\text{m}$ microspheres; with projection approximation~\cite{jones1941new}, retardance $\delta_{\phi}$ for thickness $d$ and birefringence $\delta_n$ are related as $\delta_{\phi}=2\pi nd/\lambda$)}.  
\subsection{Detecting cardiac amyloidosis from heart tissue sample}
Finally, we applied our method to image a thinly-sliced heart sample commonly viewed under polarized light microscope in pathology lab. In current practices, the cardiac tissue is thinly sectioned and stained with congo red dye. Under white light illumination, if we sandwich the sample in between a pair of cross-polarized generator and analyzer, the birefringence of amyloid, an indicator of a deadly disease called cardiac amyloidosis, can create a vibrant apple-green color~\cite{yakupova2019congo}, as shown in Fig.\ref{fig4::congored}(b). To enlarge the field-of-view, we switch to a 10x objective for this sample, and use 1.0\textmu{m} step size. Figure \ref{fig4::congored}(a) shows a brightfield image when the sample in focus. Figure \ref{fig4::congored}(c)-(d) shows the reconstructions of refractive index and birefringence. As the sample is thinly sliced, we only present a center slice of the sample. We can clearly see that the birefringence structure in (d) is very similar to (b), which can be used to as a tool to help diagnosis amyloidosis. (e) and (f) display reconstructed slices at different depth positions for the region labeled in (c) and (d).

\section{Conclusion and Discussion}
In summary, we have introduced T$^2$DPC, a new method to record polarized measurements with computational illumination strategies, and provide quantitative tomographic permittivity matrix reconstructions for a variety of calibration samples and biological specimens  {without an interferometry setup. Using relatively low NA objectives, we demonstrate by volumetric modeling the sample, T$^2$DPC has superior depth section ability compared with 2D methods.} To extend the functions of T$^2$DPC, several incremental improvements can be made. First, to relieve the ill-poseness of the problem, we have approximated the permittivity tensor with its lateral components. Although widely used~\cite{saba2021polarization,song2020ptychography,song2021large,dai2022quantitative}, this simplification disregards out-of-plane anisotropy, which can contain valuable biological and diagnostic information~\cite{yeh2021upti,yang2018polarized}. Using more angular and polarization-diverse patterns can potentially allow reconstruction of uni-axial, or even more general permittivity tensors, from which out-of-plane orientations can be derived. This can be practically implemented by placing different generators in front of each individual LEDs and turn on them sequentially. In addition, deploying measurement strategies from diffraction tomography-type setups~\cite{li2019high,zhou2020diffraction} can also hopefully record enough informative measurements for tomographic reconstructions without scanning the sample~\cite{van2020polarization,saba2021polarization}. {Further, adopting a vectorial multi-scattering forward model~\cite{saba2021polarization} can hopefully extend the method to image thicker samples.}

Finally, as a first demonstration, currently we use a stochastic gradient-based method to solve the inverse problem. While it is easy to implement, it can also result in slow convergence. As the propagation through imaging system part of the forward model can be efficiently inverted~\cite{chen20163d}, we anticipate deploying variable splitting methods~\cite{boyd2011distributed} can significantly improve the convergence speed, as well as allows flexible use of regularization methods may not have an explicit form~\cite{sun2019regularized}, for example. We hope to explore in these directions to ensure successful translation of T$^2$DPC to future clinical and scientific studies.
\section{Conflict of interest}
S.X., X.D., and R.H. have submitted a patent application related to this work, assigned to Duke University.

\ifpeerreview \else
\section*{Acknowledgments}
The authors would like to thank Dr. Shalin Mehta, Dr. Ian Sigal, Dr. Li-Hao Yeh helpful guidance and feedback. The authors would also like to thank to a Duke-Coulter Translational Partnership and funding from a 3M Nontenured Faculty Award.
\fi

\bibliographystyle{IEEEtran}
\bibliography{references}

\ifpeerreview \else

 \begin{IEEEbiography}[{\includegraphics[width=1in,height=1.25in,clip,keepaspectratio]{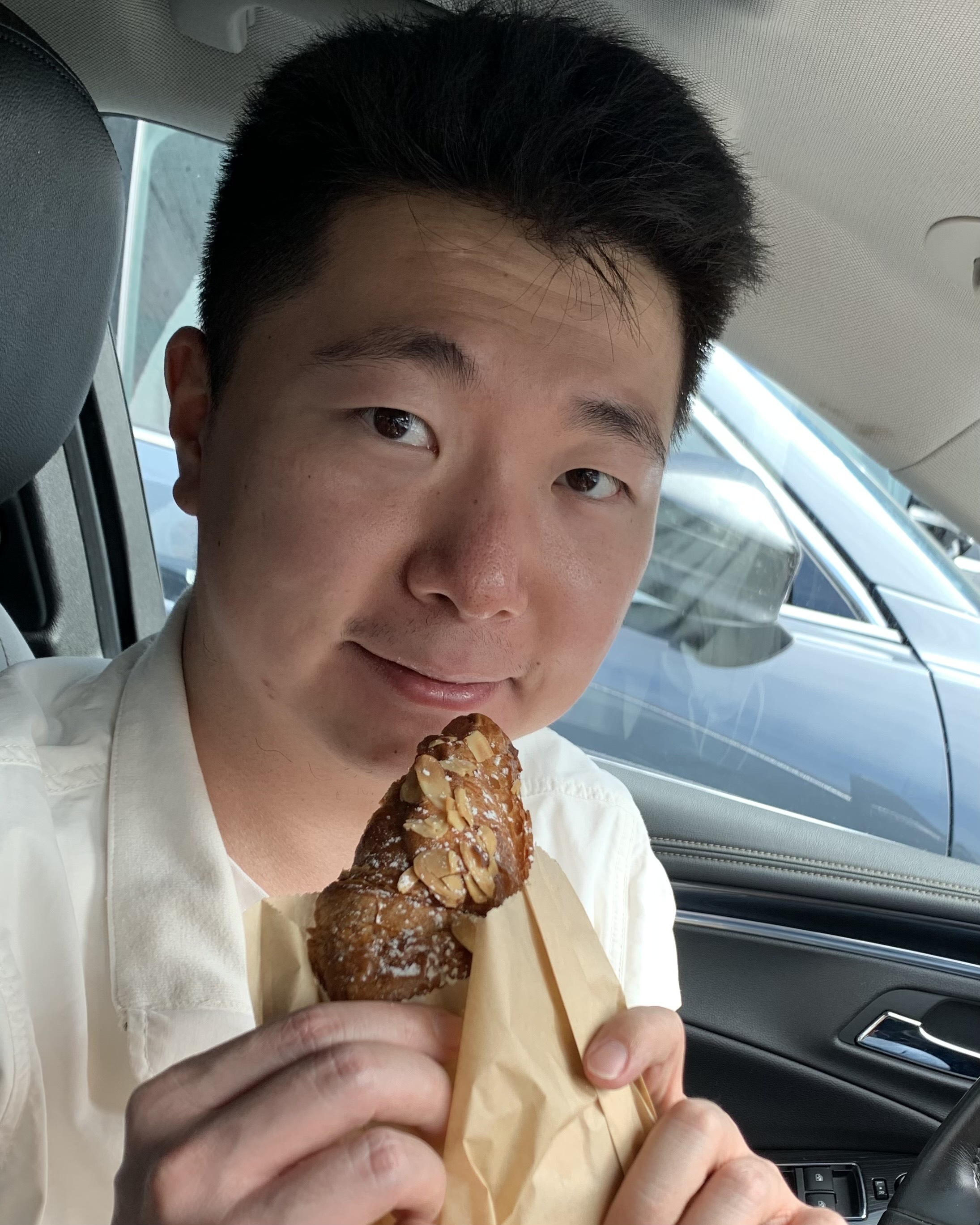}}]{Shiqi Xu}
is a Ph.D. student at Duke University. He develops computational algorithms and optical systems to image biomedical events. Most of his works are at intersections of computational imaging, applied machine learning, and optics. His current research effort falls into two major categories: non-invasive imaging deep inside living tissue and high-throughput high-content gigapixel microscopies. Before arriving at Duke, he was an M.S. student at Washington University working on computational imaging and large-scale optimization problems. Outside of research, he plays the violin and he likes David Nadien. 
\end{IEEEbiography}
\begin{IEEEbiography}[{\includegraphics[width=1in,height=1.25in,clip,keepaspectratio]{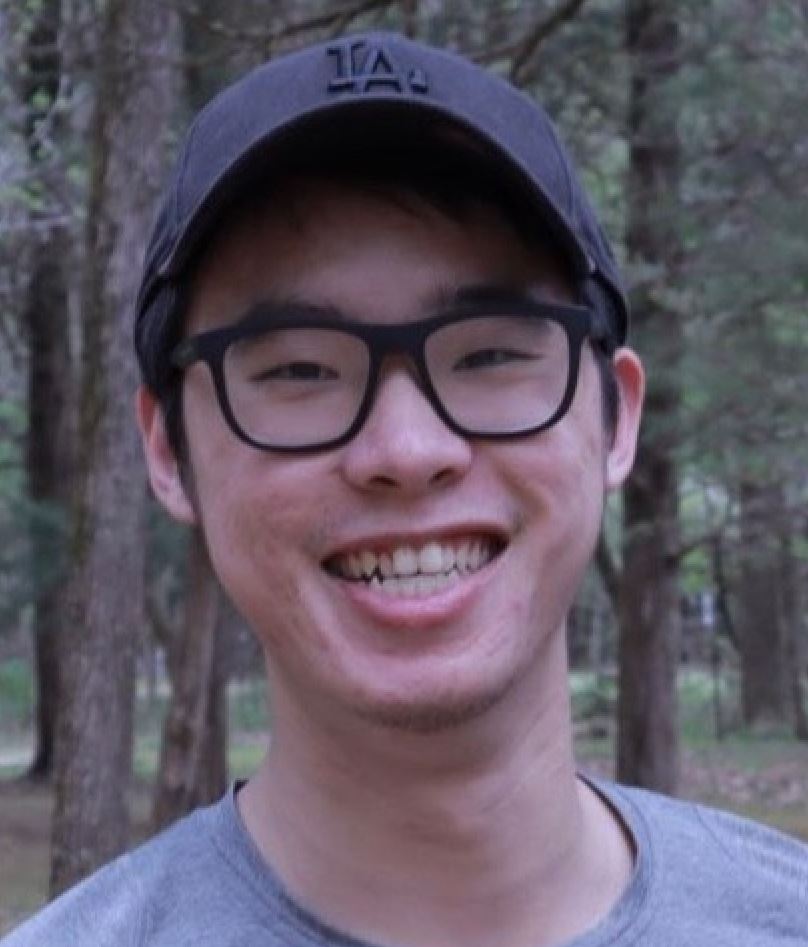}}]{Xiang Dai}
is a Ph.D. student at University of California San Diego working on computational imaging, computational photography, neuroimaging, ocean microscopy. Before arriving at UCSD, he was a M.S. student at Duke studying computational microscopy. Outside research, he enjoys playing basketball, fishing and surfing.
\end{IEEEbiography}
\begin{IEEEbiography}[{\includegraphics[width=1in,height=1.25in,clip,keepaspectratio]{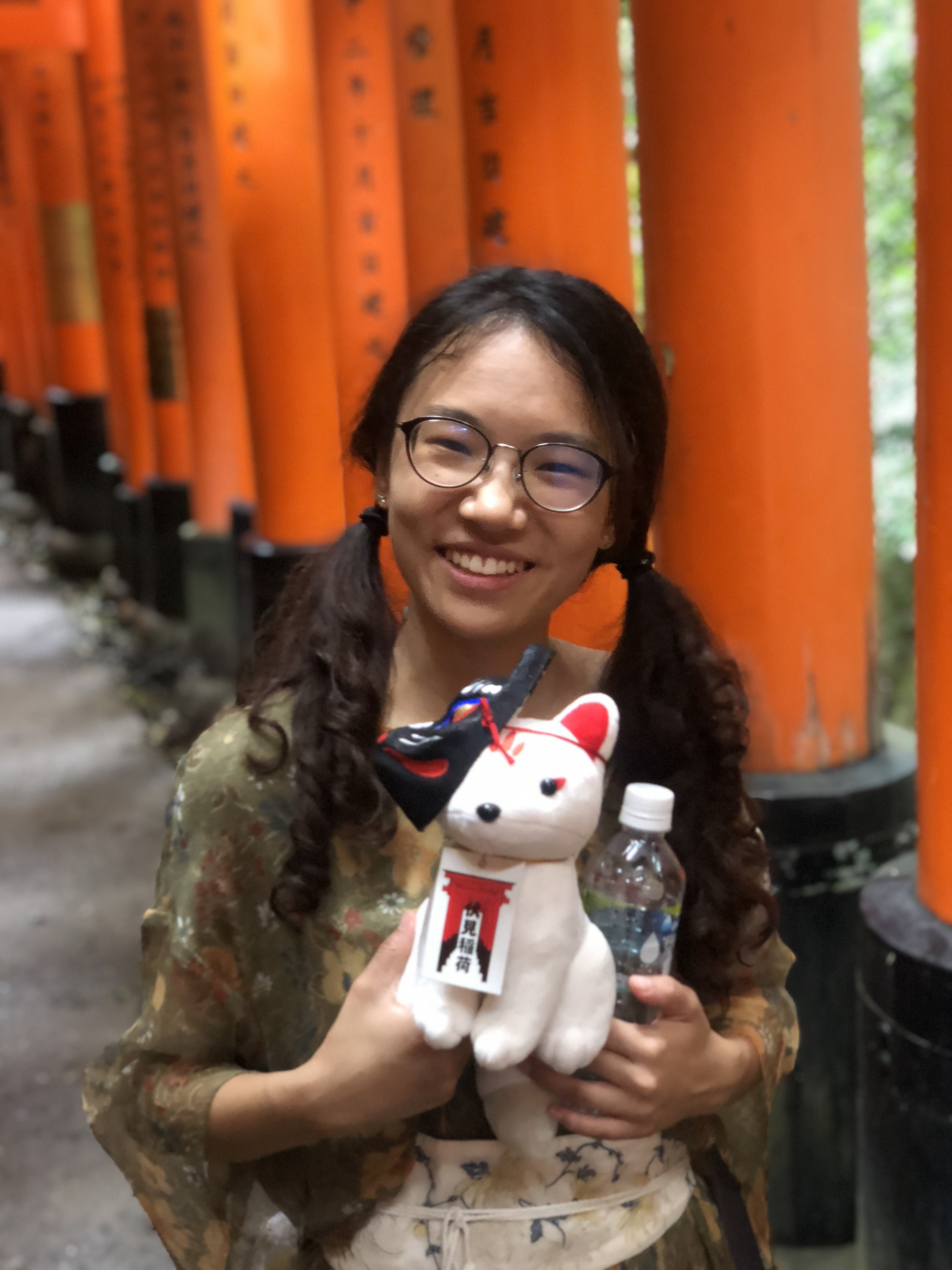}}]{Xi Yang}
is a PhD student in BME department at Duke university. She received a B.S. from Nankai University, majoring in Physics, while she also spent her high school years in Nankai High school. She has worked on self-accelerating beam and two-photon microscopy projects during her undergraduate research. Now she is helping to develop the new generation of Fourier Ptychography Microscopy in Dr. Horstmeyer’s group, after completing her Masters degree in BME at Duke over the past two years.
\end{IEEEbiography}
\begin{IEEEbiography}[{\includegraphics[width=1in,height=1.25in,clip,keepaspectratio]{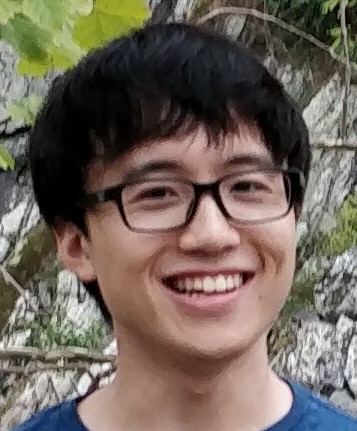}}]{Kevin Zhou}'s research interests are broadly in computational imaging, coherent (and incoherent) optical imaging, tomographic reconstruction algorithms, inverse problems, and machine learning. He has particular experience with optical coherence tomography (OCT), Fourier ptychography, diffraction tomography, and nonlinear microscopy, but he's always open to exploring and applying computational optimization techniques to other forms of imaging!
\end{IEEEbiography}
\begin{IEEEbiography}[{\includegraphics[width=1in,height=1.25in,clip,keepaspectratio]{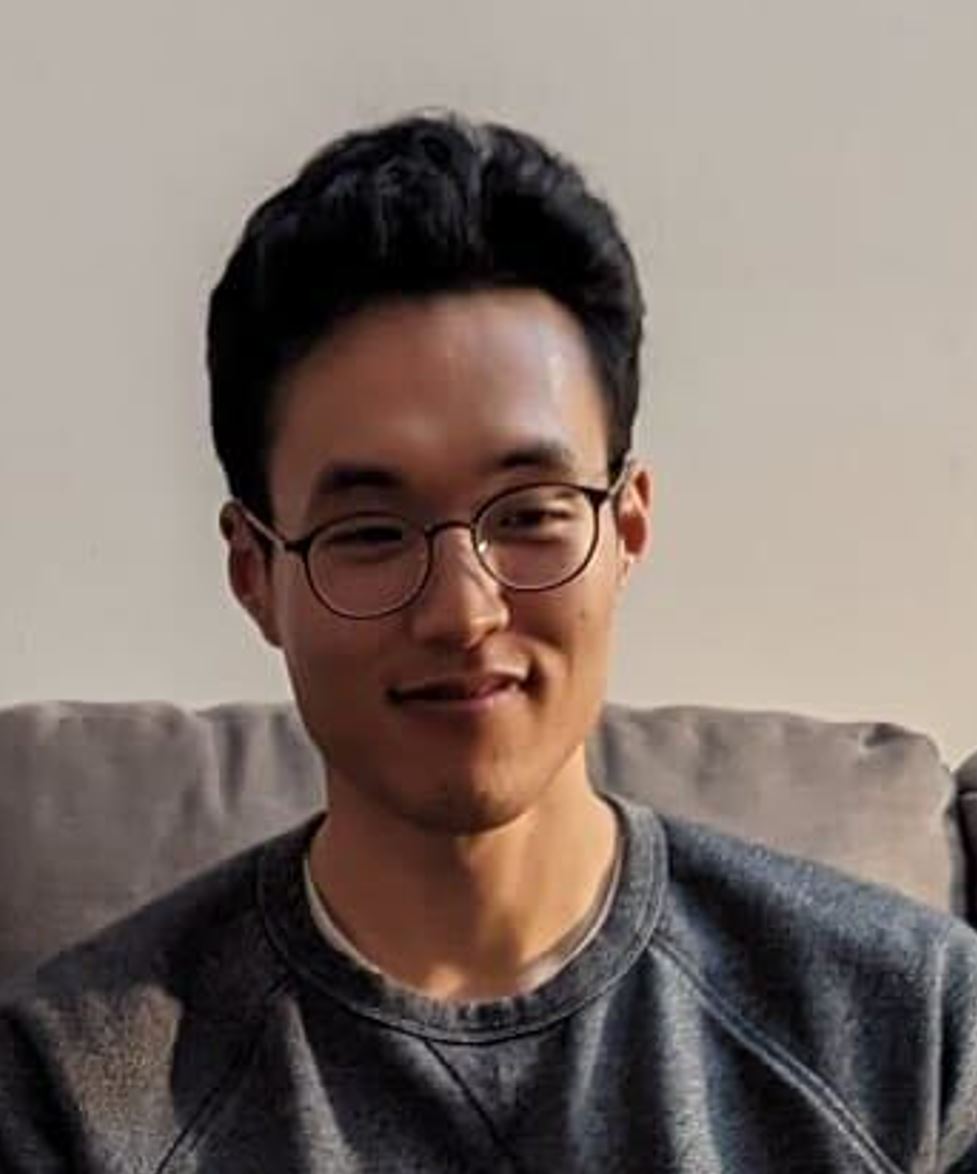}}]{Kanghyun Kim} is currently pursuing PhD in Biomedical Engineering. He recently completed an MS degree in Electrical and Computer Engineering at Duke. He received his B.S. in Statistics and a minor in Computer Science from Chung-Ang University. His research focuses on designing task-specific microscopes using a deep neural network to improve image classification accuracy. He likes to travel and take pictures. Especially, he really likes to stay in one place and watch the landscape change with the sunlight.
\end{IEEEbiography}
\begin{IEEEbiography}[{\includegraphics[width=1in,height=1.25in,clip,keepaspectratio]{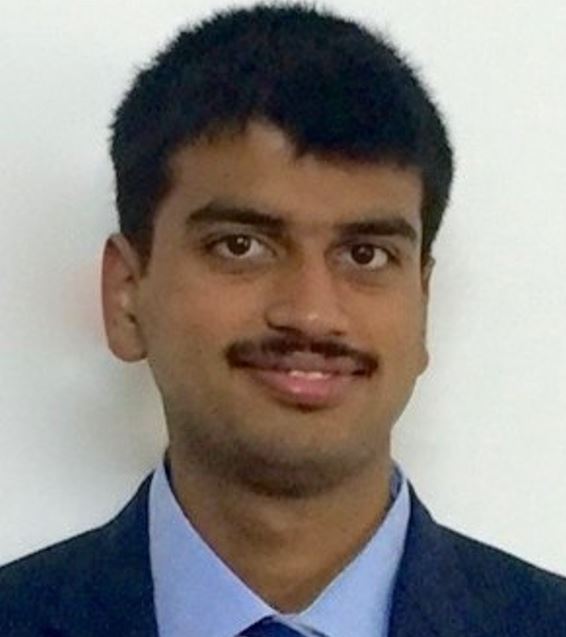}}]{Vinayak Pathak} is an aspiring biomedical engineer who is interested at the intersection of machine learning, optics and computational imaging, to improve medical imaging and diagnostics. Currently he is building miniature camera arrays for imaging mouse brain \textit{in vivo}.
\end{IEEEbiography}
\begin{IEEEbiography}[{\includegraphics[width=1in,height=1.25in,clip,keepaspectratio]{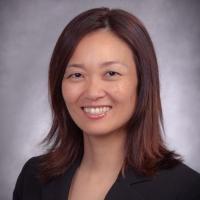}}]{Carolyn Glass} received her BS in Neuroscience with Departmental Honors from the University of California at Los Angeles (UCLA) in 1997, before earning her MS from Baylor College of Medicine and MD from the University of Texas Medical Branch, Magna Cum Laude in 2007. Dr. Glass initially trained as a vascular surgeon with a focus on endovascular/interventional procedures through the Integrated Vascular Surgery Program at the University of Rochester Medical Center.  As a recipient of the NIH National Lung Blood Institute T32 Ruth Kirschstein National Service Research Award, she completed a Ph.D with a concentration in Genomics and Epigenetics in 2014. Dr. Glass completed residency in Anatomic Pathology at the Brigham and Women’s Hospital/Harvard Medical School in 2016 followed by fellowships in Cardiothoracic Pathology also at Brigham and Women’s Hospital and Pulmonary/Transplant Pathology at the University of Texas Southwestern Medical Center. As a thoracic pathologist, Dr. Glass also has a special interest in identifying new epigenetic biomarkers that may predict response or resistance to conventional, targeted and immune therapy using computational techniques. She works closely with the Duke Thoracic Oncology Group and DCI Center for Cancer Immunotherapy. 
\end{IEEEbiography}
\begin{IEEEbiography}[{\includegraphics[width=1in,height=1.25in,clip,keepaspectratio]{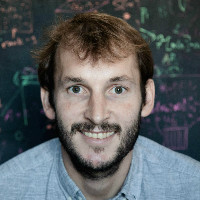}}]{Roarke Horstmeyer} is an assistant professor within Duke's Biomedical Engineering Department. He develops microscopes, cameras and computer algorithms for a wide range of applications, from forming 3D reconstructions of organisms to detecting neural activity deep within tissue. His areas of interest include optics, signal processing, optimization and neuroscience. Most recently, Dr. Horstmeyer was a guest professor at the University of Erlangen in Germany and an Einstein postdoctoral fellow at Charitè Medical School in Berlin. Prior to his time in Germany, Dr. Horstmeyer earned a PhD from Caltech’s electrical engineering department in 2016, a master of science degree from the MIT Media Lab in 2011, and a bachelors degree in physics and Japanese from Duke University in 2006.
\end{IEEEbiography}




\fi

\end{document}